\documentclass{article}
\usepackage[numbers]{natbib}
\usepackage{amsmath}
\usepackage{graphicx}
\usepackage[english]{babel}
\usepackage{authblk}
\usepackage[a4paper, margin=1in]{geometry}
\begin{document}

%%%% Article title to be placed here
\title{Variable-order fractional master equation and clustering of particles: non-uniform lysosome distribution}

\author[1]{Sergei Fedotov}
\author[1,2]{Daniel Han}
\author[3]{Andrey Yu. Zubarev}
\author[2]{Mark Johnston}
\author[2]{Victoria J Allan}
\affil[1]{Department of Mathematics, University of Manchester, M13 9PL}
\affil[2]{Faculty of Biology, Medicine and Health, School of Biological Sciences, University of Manchester, M13 9PL}
\affil[3]{Ural Federal University, 620083 Ekaterinburg, Russia}

%%%% Subject entries to be placed here %%%%
%\subject{xxxxx, xxxxx, xxxx}

%%%% Keyword entries to be placed here %%%%
%\keywords{xxxx, xxxx, xxxx}

%%%% Insert corresponding author and its email address}
%\corres{Insert corresponding author name\\
%\email{xxx@xxxx.xx.xx}}

%%%% Abstract text to be placed here %%%%%%%%%%%%

%%%%%%%%%%%%%%%%%%%%%%%%%%%

%%%%%%%%%% Insert the texts which can accomdate on firstpage in the tag "fmtext" %%%%%

%\begin{fmtext}
%\end{fmtext}
%%%%%%%%%%%%%%% End of first page %%%%%%%%%%%%%%%%%%%%%

\maketitle

\begin{abstract}
	In this paper, we formulate the space-dependent variable-order fractional master equation to model clustering of particles,  organelles, inside living cells. We find its solution in the long time limit describing non-uniform distribution due to a space dependent fractional exponent. In the continuous space limit, the solution of this fractional master equation is found to be exactly the same as the space-dependent variable-order fractional diffusion equation. In addition, we show that the clustering of lysosomes, an essential organelle for healthy functioning of mammalian cells, exhibit space-dependent fractional exponents.  Furthermore, we demonstrate that the non-uniform distribution of lysosomes in living cells is accurately described by the asymptotic solution of the space-dependent variable-order fractional master equation. Finally, Monte Carlo simulations of the fractional master equation validate our analytical solution.
\end{abstract}

\section{Introduction}
%%%% Insert A head here

Anomalous transport has gained much interest due to its applications in physics, chemistry, biology \cite{metzler2000random,klages2008anomalous,mendez2010reaction,klafter2011first}. It has proven to be a powerful theory to characterise dynamics of biological processes using the fractional exponent, $\mu$ for the mean squared displacement, $\langle x^2(t) \rangle \propto t^{\mu}$. When describing the broad ensemble statistics of random walkers, a constant fractional exponent suffices. For cell biology \cite{yuste2010reaction,sokolov2012models,burov2011single,krapf2015mechanisms,angstmann2016mathematical}, studies of tracer particles in mammalian cell cytoplasm \cite{weiss2004anomalous}, \textit{in cellulo} vesicle transport \cite{chen2015memoryless,fedotov2018memory,woringer2020anomalous} and loci in bacteria \cite{yu2018subdiffusion} were all shown to be well described through a constant fractional exponent. Quantifying dynamic cellular processes has been a major success for anomalous transport theory and much scientific work is still ongoing \cite{song2018neuronal,berry2014spatial,kalwarczyk2017apparent,munoz2020single,charalambous2017nonergodic,jeon2016protein}. However, current experimental studies \cite{han2020deciphering,garamella2020anomalous,sabri2020elucidating} are finding evidence of heterogeneous anomalous transport in intracellular processes while the theory for heterogeneous anomalous transport (specifically when the fractional exponent $\mu$ is no longer a constant) remains largely neglected. In fact, it is given knowledge that the cellular cytoplasm is a vastly heterogeneous complex fluid \cite{alberts2018molecular}.

In particular, organelles responsible for cellular metabolism and degradation called lysosomes move predominantly subdiffusively with heterogeneous fractional exponents that depend on their spatial positioning \cite{ba2018whole,han2020deciphering}. This implies that lysosome dynamics should be adequately described by a fractional diffusion equation with a space dependent fractional exponent, $\mu(x)$. More interestingly, lysosomes also maintain a stable non-uniform spatial pattern clustered near the centrosome in the cell \cite{ba2018whole}. So a challenge for modelling was posed: What is the asymptotic distribution to the space-dependent variable-order fractional diffusion equation \cite{chechkin2005fractional} and does the experimental distribution of lysosomes match this asymptotic distribution? 

Recently, we found the asymptotic representation of the solution of the variable-order fractional diffusion equation analytically with an ultra-slow spatial clustering (aggregation) of subdiffusive particles \cite{fedotov2019asymptotic}. This equation is 
\begin{align}
\frac{\partial p(x,t)}{\partial t} = \frac{\partial^2}{\partial x^2} \left[ D_{\mu(x)} \mathcal{D}_t^{1-\mu(x)}p(x,t) \right]
\label{vofde}
\end{align}
where $p(x,t)$ is the PDF of a particle at position $x$ and time $t$. The function $p(x,t)$ can also be interpreted as the mean density of particles at $x$ and $t$; $\mu(x)$ is the space dependent anomalous exponent; $D_{\mu(x)} = a^2/2\tau_0^{\mu(x)}$ is the fractional diffusion coefficient with a time scale $\tau_0$ and length scale $a$; and $\mathcal{D}_t^{1-\mu(x)}$ is the Riemann-Liouville derivative. Since we are interested heterogeneous subdiffusion, $\mu(x)\in(0,1)$. The Riemann-Liouville derivative is defined as
\begin{align}
\mathcal{D}_t^{1-\mu(x)}p(x,t) = \frac{1}{\Gamma(\mu(x))}\frac{\partial}{\partial t} \int_{0}^{t}\frac{p(x,t')}{(t-t')^{1-\mu(x)}}dt'
\notag
\end{align}
For a monotonically increasing fractional exponent with domain $x\in[0,L]$, the asymptotic solution to \eqref{vofde} is \cite{fedotov2019asymptotic}
\begin{align}
p(x,t) \sim \frac{\mu_0'\left(\frac{t}{\tau_0}\right)^{\Delta \mu(x)}}{\Gamma(1-\Delta \mu(x))}\left[\ln\left(\frac{t}{\tau_0}\right)-\psi_0(1-\Delta\mu(x))\right]
\label{asympsol}
\end{align}
where $\Delta \mu(x) = \mu(x)-\mu(0)$, $\mu_0' = \frac{d\mu}{dx}(0)\neq0$ and $\psi_0(x)$ is the digamma function. In \cite{fedotov2019asymptotic}, we solved \eqref{vofde} directly using the Laplace transform in the long time limit. The asymptotic solution in Laplace space corresponding to \eqref{asympsol} is
\begin{align}
	s\hat{p}(s,t) \sim -\left(\tau_0 s\right)^{\Delta \mu(x)} \mu_0' \ln(\tau_0 s).
	\label{laplace_asympsol}
\end{align}
The PDF in \eqref{asympsol} describes a non-uniform distribution clustered about the point of minimum $\mu(x)$ located at $x=0$. This clustering has no analogue in the classical advection-diffusion models. 

In this paper, we formulate the space-dependent variable-order fractional master equation. The purpose is to show that the solution converges to that of the space-dependent variable-order fractional diffusion equation presented in \cite{fedotov2019asymptotic}. Furthermore, we demonstrate that the probability density function (PDF) for the asymptotic representation of the solution of the variable-order fractional diffusion equation matches the observed density of lysosomes clustering near the centrosome in living cells \cite{ba2018whole}. In what follows, we will demonstrate that the same solution is obtained from the continuous approximation to the discrete master equation that generates \eqref{vofde}.

\section{Space-dependent variable-order fractional master equation}
\begin{figure}[h!]
	\centering
	\includegraphics[width=0.35\linewidth]{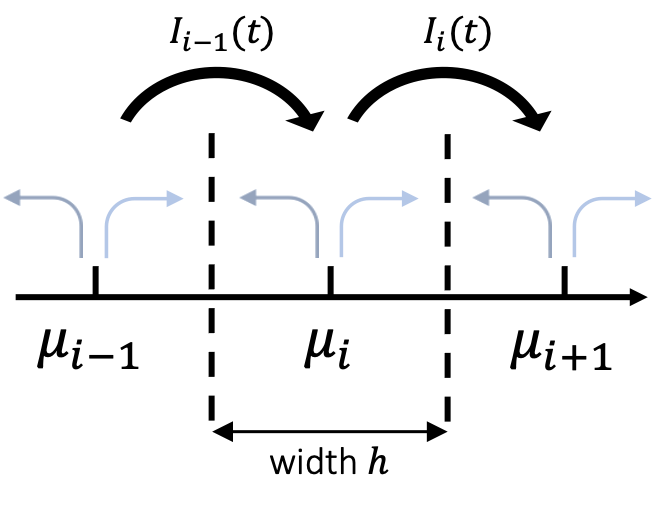}
	\caption{A diagram that shows the subinterval $i$ with boundaries $[(i-1)h,ih]$ drawn with dashed lines labelled with the corresponding fractional exponent $\mu_i$. The subintervals on either side have fractional exponent $\mu_{i-1}$ and $\mu_{i+1}$ and every subinterval has width $h$. The small angled arrows represent particles leaving each subinterval and the large round arrows show the flux of particles for subinterval $i$.}
	\label{diagram}
\end{figure}

In this section, our aims are to formulate a master equation for the subdiffusive movement of particles in heterogeneous media and find the solution of this master equation. Consider the random movement of a particle in a domain $[0,L]$ and divide it into $N$ subintervals of length $h = \frac{L}{N}$. The mean number of particles in the subinterval $i$, spanning $[(i-1)h,ih]$, is denoted as $n_i(t)$ for $1\leq i \leq N$. Every subinterval, $i$, is characterized by a fractional exponent, $\mu_i$. The non-homogeneous fractional master equation can be written as \cite{fedotov2013non,fedotov2013random}
\begin{align}
\begin{split}
\frac{dn_i(t)}{dt} =& -I_{i}(t) + I_{i-1}(t) \\ =& \frac{1}{2\tau_0^{\mu_{i-1}}} \mathcal{D}^{1-\mu_{i-1}}_{t}n_{i-1}(t) + \frac{1}{2\tau_0^{\mu_{i+1}}}\mathcal{D}^{1-\mu_{i+1}}_{t}n_{i+1}(t) - \frac{1}{\tau_0^{\mu_{i}}} \mathcal{D}^{1-\mu_{i}}_{t}n_{i}(t) 
\end{split}
\label{master}
\end{align}
where the fluxes, 
\begin{align}
\begin{split}
I_{i}(t) &= \frac{1}{2\tau_0^{\mu_{i}}}\mathcal{D}_t^{1-\mu_{i}}n_{i}(t) - \frac{1}{2\tau_0^{\mu_{i+1}}}\mathcal{D}_t^{1-\mu_{i+1}}n_{i+1}(t) \\
I_{i-1}(t) &= \frac{1}{2\tau_0^{\mu_{i-1}}}\mathcal{D}_t^{1-\mu_{i-1}}n_{i-1}(t) - \frac{1}{2\tau_0^{\mu_{i}}}\mathcal{D}_t^{1-\mu_{i}}n_{i}(t).
\end{split}
\label{fluxdefinition}
\end{align}
The flux of particles from subinterval $i$ to $i+1$ is $I_i(t)$ and from subinterval $i-1$ to $i$ is $I_{i-1}(t)$. A schematic of \eqref{master} showing the fluxes is seen in Figure \ref{diagram}

Since we assume there is no external flux of particles entering our domain $[0.L]$, the total mass is conserved:
\begin{align}
\sum_{i=1}^{N}n_i(t) = n.
\label{conservation}
\end{align}
For the boundary intevals, $i=1$ and $i=N$ in \eqref{master}, we have
\begin{align}
\frac{dn_i(t)}{dt} = -I_1(t), \hspace{0.1cm} \frac{dn_N(t)}{dt} = I_{N-1}(t).
\label{1Nmaster}
\end{align}
Taking the Laplace transform of \eqref{master}, we obtain
\begin{align}
s\hat{n}_i(s) - n_i(0) = -\hat{I}_i(s) + \hat{I}_{i-1}(s)
\label{masterfluxeslaplace}
\end{align}
for $2\leq n \leq N-1$, where Laplace transforms of \eqref{fluxdefinition} are
\begin{align}
\begin{split}
\hat{I}_{i}(s) &= \frac{s\hat{n}_{i}(s)}{2(\tau_0 s)^{\mu_{i}}} -  \frac{s\hat{n}_{i+1}(s)}{2(\tau_0 s)^{\mu_{i+1}}},\\
\hat{I}_{i-1}(s) &= \frac{s\hat{n}_{i-1}(s)}{2(\tau_0 s)^{\mu_{i-1}}} -  \frac{s\hat{n}_{i}(s)}{2(\tau_0 s)^{\mu_{i}}}.
\end{split}
\label{fluxdefinitionlaplace}
\end{align}
For $i=1$ and $i=N$, \eqref{1Nmaster} becomes
\begin{align}
\begin{split}
s\hat{n}_1(s) - n_1(0) &= -\hat{I}_1(s)\\
s\hat{n}_N(s) - n_N(0) &= \hat{I}_{N-1}(s).
\end{split}
\label{1Nmasterfluxeslaplace}
\end{align}
In the long time limit, $\tau_0s \rightarrow 0$, since $\hat{I}_{i-1}(s)\approx0$ in \eqref{fluxdefinitionlaplace}, one can obtain $\hat{n}_i(s)$ in terms of $\hat{n}_{i-1}(s)$ as
\begin{align}
\hat{n}_i(s) \simeq (\tau_0 s)^{\mu_i-\mu_{i-1}}\hat{n}_{i-1}(s)
\label{ni_nip1}
\end{align}

Now we define $\mu_i-\mu_{i-1} = \alpha h$ such that the difference between exponents at two neighboring sites tends to zero as $N\rightarrow\infty$. For $\mu(x)$ a linear function of $x$, $\alpha$ will be the gradient. Then \eqref{ni_nip1} becomes
\begin{align}
\hat{n}_i(s) \simeq (\tau_0 s)^{\alpha h}\hat{n}_{i-1}(s).
\label{ni_nip1_alpha}
\end{align}
By summing \eqref{masterfluxeslaplace} and \eqref{1Nmasterfluxeslaplace} for all $i$ and using the conservation of total mass from \eqref{conservation}, we can obtain
\begin{align}
\sum_{i=1}^{N} s\hat{n}_i(s) = n.
\label{conservation2}
\end{align}
Then using the recursive relation \eqref{ni_nip1_alpha}, \eqref{conservation2} can be written as
\begin{align*}
s\hat{n}_1(s) \left[ 1 + (\tau_0s)^{\alpha h} + (\tau_0s)^{2\alpha h}  \cdots + (\tau_0s)^{(N-1)\alpha h} \right]= n.
\end{align*}
So, we can find the solution of $\hat{n}_1(s)$ in Laplace space as
\begin{align}
s\hat{n}_1(s) & = \frac{n}{\sum_{k=0}^{N-1}(\tau_0s)^{k\alpha h}} = n\frac{1-\left(\tau_0s\right)^{\alpha h}}{1-\left(\tau_0 s\right)^{N\alpha h}}.
\label{n1sol}
\end{align}
Using \eqref{ni_nip1_alpha} together with \eqref{n1sol}, we find the solution of  $\hat{n}_i(s)$, for $2\leq i \leq N$, in Laplace space as
\begin{align}
s\hat{n}_i(s) = n\left(\tau_0 s\right)^{(i-1)\alpha h } \frac{1-\left(\tau_0s\right)^{\alpha h}}{1-\left(\tau_0 s\right)^{N\alpha h}}.
\label{nisol}
\end{align} 

For the asymptotic limit $\tau_0s \rightarrow 0$, \eqref{n1sol} and \eqref{nisol} become
\begin{align}
\begin{split}
s\hat{n}_1(s) &\simeq n (1-\left(\tau_0s\right)^{\alpha h}),\\
s\hat{n}_i (s) &\simeq n (\tau_0s)^{(i-1)\alpha h}(1-\left(\tau_0s\right)^{\alpha h}).
\end{split}
\label{asympsollaplace}
\end{align}
To verify that the master equation \eqref{master} does indeed correspond to the space-dependent variable-order fractional diffusion equation in \cite{fedotov2019asymptotic}, we need to introduce the density 
\begin{align}
\begin{split}
\hat{p}_i(s) &= \frac{\hat{n}_i(s)}{h}, \text{ then}\\
s\hat{p}_i(s) &= n \frac{(\tau_0s)^{(i-1)\alpha h}(1-\left(\tau_0s\right)^{\alpha h})}{h}.
\end{split}
\end{align}
Then setting $x=ih$ and using the well known formula
\begin{align*}
\lim\limits_{h\rightarrow0} \frac{(\tau_0s)^h-1}{h} = \alpha\ln(\tau_0s),
\end{align*}
we obtain, in the continuous limit,
\begin{align}
s\hat{p}(x,s) = \lim\limits_{h\rightarrow 0}s\hat{p}_i(t) =  -\alpha n (\tau_0s)^{x\alpha} \ln(\tau_0s).
\label{laplace_asympsol2}
\end{align}
This is equivalent to the solution found in Ref. \cite{fedotov2019asymptotic}, shown as \eqref{laplace_asympsol} in this paper. Monte Carlo simulations for the master equation \eqref{master} were performed and shown to have excellent correspondence with this asymptotic solution as shown in Figure \ref{fig:lysodata}. The simulations were made by generating random residence times, $T$, for particles in box $i$ drawn from the PDF $\psi_{\mu_i}(\tau) = -(\partial / \partial \tau) E_{\mu_i}[-(\tau/\tau_0)^{\mu_i}]$ (details can be found in \cite{fulger2008monte}) after which the particle jumps right or left with equal probability. The fractional exponent $\mu_i$ was increasing linearly as $i$ increased (full details can be found in \cite{fedotov2019asymptotic}). In the next section, we show that this asymptotic solution \eqref{laplace_asympsol2} and \eqref{asympsol} corresponds to the distribution of lysosomes in living cells.

\section{Experimental Evidence}
\begin{figure}[h!]
	\centering
	\includegraphics[width=0.5\linewidth]{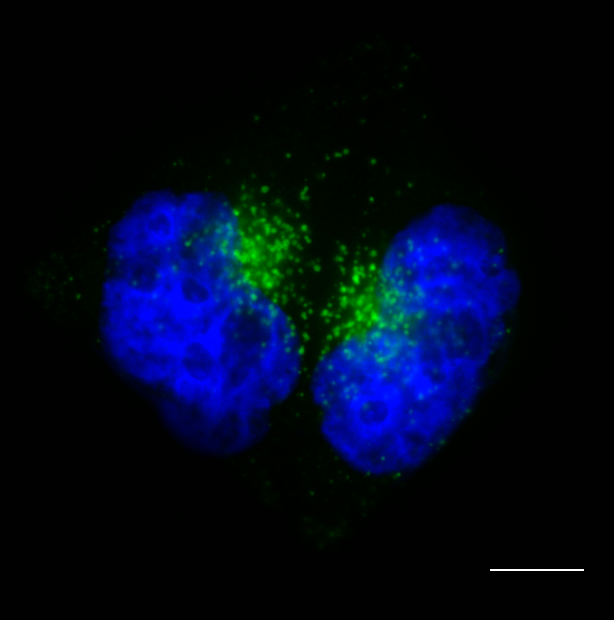}
	\caption{Non-uniform lysosome distribution in two HeLaM cells. Methanol fixed HeLaM cells were labeled with antibodies to the lysosomal protein LAMP1 (Lysosomes, green) and DAPI to label DNA in the nucleus (blue).  Lysosomes are non-uniformly distributed with a large cluster around the perinuclear region and fewer lysosomes throughout the rest of the cell. Scale bar shows $10$$\mu$m.}
	\label{fig:lysocell}
\end{figure}

Lysosomes are intracellular organelles that degrade macromolecules and regulate metabolism \cite{lawrence2019lysosome}. It is well-established that under normal conditions, the majority of them are concentrated in the perinuclear area, although at any one time a fraction of them are undergoing active bi-directional movement towards and away from the nucleus \cite{ba2018whole}. Figure \ref{fig:lysocell} shows the non-uniform distribution of lysosomes (green).
%use Ba 2018 to say that the positioning of lysosomes is not just based on subdiffusion so the next sentence needs to change.
However, the exact mechanisms for how lysosomes maintain such a macroscopic spatial distribution remain unclear. The aim of this subsection is to show that lysosomal distributions in the cell can be explained to a large extent by the anomalous mechanism detailed in this paper, since subdiffusion is the most prevalent characteristic in lysosomal movement \cite{ba2018whole,han2020deciphering}. Anomalous subdiffusion can occur as a result of non-uniform crowdedness \cite{weiss2004anomalous} in the cytoplasm. Our hypothesis is that the non-uniform fractional exponent $\mu(x)$ can serve as a measure of crowdedness in the cytoplasm such that $x$ is the distance away from the nucleus and that lysosomes display a stable distribution across cells and times due to this non-uniform subdiffusion.

\begin{figure}[h!]
	\centering
	\includegraphics[width=\linewidth]{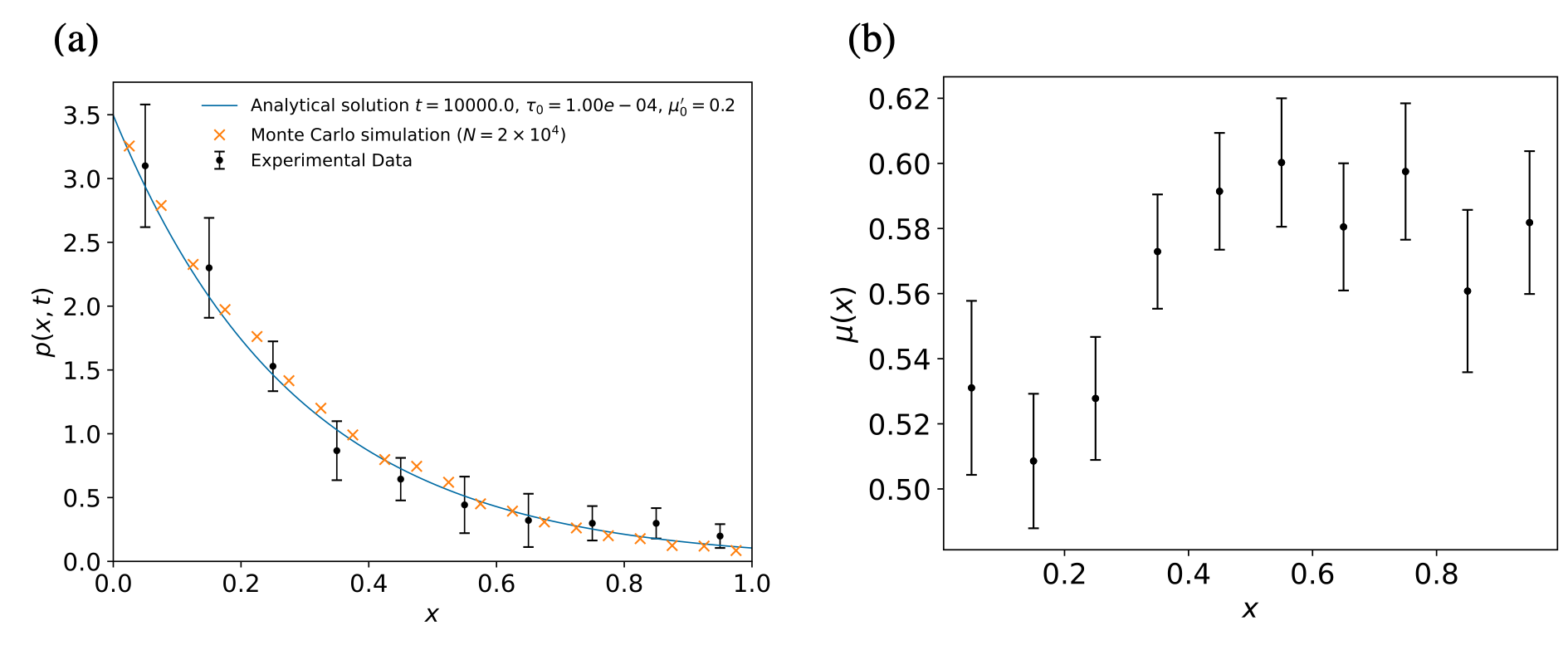}
	\caption{ Data from live-cell imaging experiments of HeLaM cells with LysoBrite labelled lysosomes. (a) Normalized density of lysosomes, $p(x,t)$, against the normalized displacement away from the cell centre inside 30 HeLaM cells. The experimental data (black dots) and fit of the asymptotic solution (blue solid line) show excellent correspondence with parameters $\mu_0’ = 0.149$, $\tau_0 = 9.433\times10^{-8}$ and $t = 8.829 \times 10^2$. The Monte Carlo simulations  (orange crosses) of $N=2\times 10^4$ particles with the same parameters as the fit also show excellent correspondence with the analytical solution. (b) The plot of experimentally measured fractional exponent $\mu(x)$ against the normalized displacement away from the cell centre for 30 HeLaM cells.}
	\label{fig:lysodata}
\end{figure}

We performed live-cell imaging experiments to analyze lysosome positions; experimental and analysis methods are detailed in Section \ref{experimentalmethods}. Figure \ref{fig:lysodata}a shows the empirical PDF (points) of finding a lysosome at a certain distance from the cell center from a sample of HeLaM cells and the asymptotic PDF \eqref{asympsol} (line) for parameters $\mu'_0 =  0.149 $, $\tau_0 = 9.433\times10^{-8}$s and $t = 8.829\times10^{2}$s.
One can see that the prediction corresponds well to the empirical PDF.
Furthermore, we measured the fractional exponents directly from the same HeLaM cells by calculating the time-averaged mean squared displacement $\langle r^2(\tau)\rangle$ of each trajectory and fitting to a power law $\tau^{\mu}$. Then from the lysosome population with $0<\mu<1$ and track lengths of greater than 100 points (1725 trajectories from 30 cells), the normalized average displacement from the cell centre was calculated. The plot of these data is shown in Figure \ref{fig:lysodata}b Most importantly, it shows that the fractional exponent is indeed space-dependent and in general increases from $x=0$ to $x=1$. The fluctuations in $\mu(x)$ near $x=0$ and $x=1$ is due to low number of lysosomes found exactly at the cell centre or the cell periphery.

%\begin{figure}[h!]
%	\centering
%	\includegraphics[width=0.5\linewidth]{dist_avg_cells.eps}
%	\caption{Normalized density of lysosomes, $p(x,t)$, against the normalized displacement away from the cell centre inside 30 HeLaM cells. The experimental data (black dots) and fit of the asymptotic solution (blue solid line) show excellent correspondence with parameters $\mu_0’ = 0.149$, $\tau_0 = 9.433\times10^{-8}$ and $t = 8.829 \times 10^2$.}
%	\label{fig:lysodata}
%\end{figure}
%
%\begin{figure}[h!]
%	\centering
%	\includegraphics[width=0.5\linewidth]{anomalous_exponent_function.eps}
%	\caption{The plot of measured fractional exponent $\mu(x)$ against the normalized displacement away from the cell centre for 30 HeLaM cells.}
%	\label{fig:lysoanomexpdata}
%\end{figure}

%However, nutrient starvation causes increase in perinuclear concentration of lysosomes \cite{korolchuk2011lysosomal,li2016molecular}, due to the suppression of active transport along microtubules \cite{ba2018whole}. This implies that purely subdiffusive lysosomes would aggregate closer to the nucleus, supporting the anomalous mechanism detailed in this paper.

%Underlying subdiffusive mechanism with active transport on top
%Stress granules
%Emphasise

The anomalous mechanism presented in this paper is obviously not a complete theory to describe the non-uniform distribution of intracellular organelles. There are many other interactions and phenomena that occur in conjunction. Two primary additional phenomena that will affect this pattern is the superdiffusion generated by motor protein transport of organelles \cite{chen2015memoryless,fedotov2018memory,korabel2018non} and the non-linear interaction of subdiffusive organelles \cite{straka2015transport} such as the lysosome tethering to the endoplasmic reticulum observed in \cite{ba2018whole}. Furthermore, there are several other mechanisms, such as viscoelasticity and diffusion in labyrinthine environments, that lead to subdiffusive motion of organelles (see the excellent review \cite{sokolov2012models}). Including these additional effects in future works should provide a more physical and accurate model of organelle organization in the cell. However, \eqref{asympsol} models the long time limit of lysosomal distributions where the effects of heterogeneous subdiffusion in the cell will dominate. 

\section{Experimental methods}
\label{experimentalmethods}
HeLaM cells were maintained in Dulbecco's Modified Eagle's Medium (DMEM) - high glucose (Sigma-Aldrich, Dorset, UK) with 10\% Fetal Bovine Serum (GE Healthcare, Buckinghamshire, UK) at $37^{\circ}$C and 8\% CO$_2$.
%and split every 2-3 days or when confluent.

For fixed images like Figure \ref{fig:lysocell},
HeLaM cells were seeded onto \#1.5 glass coverslips the day before fixation. Cells were fixed in $-20^{\circ}$C methanol for six minutes, rinsed in phosphate buffered saline (PBS) and labelled with
%Growth media was removed and cells fixed in $-20^{\circ}$C Methanol (...) for 6 minutes and then washed in PBS (...). Excess PBS was removed and coverslips stained with
mouse LAMP1 primary antibody (1/500 dilution) (Product Number: H4A3, Developmental Studies Hybridoma Bank, Iowa City, USA).
%(...) for 30 minutes followed by three 5 minute washes in PBS. During all subsequent steps, coverslips were kept in the dark as much as possible. Excess PBS was again removed and the cells further stained with
After washing in PBS, the coverslips were incubated with donkey anti-mouse Alexa594 secondary antibody (1/800 dilution) (Jackson Immuno Research Laboratories Inc., West Grove, USA). Coverslips were left for 30 minutes at room temperature, washed once with PBS for 5 minutes and then stained with 4',6-diamidino-2-phenylindole (DAPI) (0.1$\mu$g mL$^{-1}$ in PBS) for 5 minutes, followed by a final
%5 minute
wash with PBS. Coverslips were then mounted onto glass slides using ProLong\textsuperscript{\textregistered} Diamond mounting agent (Life Technologies, Paisley, UK). Cells were imaged by fluorescence microscopy using an Olympus BX Microscope, using a 60$\times$/1.4 objective, CoolSNAP EZ CCD camera (Photometrics, Tucson, USA) and MetaMorph software (Molecular Devices, San Jose, USA).

For live imaging, the HeLaM cells were first transfected with EGFP-C1 (Clontech Laboratories, Inc. CA, USA GenBank Accession \#: U55763 Catalog \#: 6084-1) using Polyplus jetPEI (Polyplus-transfection SA, France) but with half the recommended amounts of reagents. The ratios for a single 35mm dish are 1000ng of DNA, 4$\mu$L of jetPEI and 100$\mu$L of NaCl (150mM). Then they were stained with LysoBrite Red (AAT Bioquest), imaged using fluorescence microscopy and the lysosomes were tracked using Imaris. The cells were grown in MEM (Sigma Life Science) and 10\% FBS (HyClone) and incubated for 48 hr at 37 in 5\% CO\textsubscript{2} on 35 mm glass-bottomed dishes ($\mu$-Dish, Ibidi, Cat. No. 81150). LysoBrite was diluted 1 in 500 with Hank's Balanced Salt solution (Sigma Life Science), then 0.5 mL of this solution was added to cells on a 35 mm dish containing 2 mL of growing media and incubated at 37 for at least 1 hr. Then, cells were washed with sterile PBS and the media replaced with growing media.

After at least 6 hr incubation at 37$^{\circ}$C in 8\% CO\textsubscript{2}, the growing media was replaced with live-imaging media composed of Hank's Balanced Salt solution (Sigma Life Science, Cat. No. H8264) with added essential and non-essential amino acids, glutamine, penicillin/streptomycin, 25 mM HEPES (pH 7.0) and 10\% FBS (HyClone). Live-cell imaging was performed on an inverted Olympus IX71 microscope with an Olympus 100$\times$ 1.35 NA oil PH3 objective. Samples were illuminated using an OptoLED (Cairn Research) light source with 470 nm and white LEDs. For Lysobrite-Red, a white light LED, Chroma ET573/35 was used with a dualband GFP/mCherry dichroic and an mCherry emission filter (ET632/60).

Immediately before acquiring image streams, a single image of the cytoplasmic fluorescent marker EGFP-C1 (transfected, see earlier) was taken so that cell boundaries could be manually segmented in ImageJ. The cell centers were calculated in Python3 by using a Gaussian filter and then finding the point of maximum intensity. From the cell boundaries, scaled contours (of 10\% increments) centered at the cell center could be calculated. Then the fluorescence intensity within each of these scaled contours was measured and the density of lysosome fluorescence was calculated subsequently by dividing the total fluorescence intensity within the scaled contour by the number of pixels contained in the scaled contour region. This data and analysis was used for Fig. \ref{fig:lysodata}. The fit of this PDF to the solution \eqref{asympsol} was then performed in Python using the scipy.optimize package. Tracking of lysosomes in experimental videos yielded 8476 tracks in total from 30 HeLaM cells and was performed using Imaris. The time-averaged mean squared displacements were calculated from the trajectories via
\begin{align}
\langle r^2(m\delta \tau) \rangle = \frac{1}{N-m}\sum_{i=1}^{N-m}\left([x(t_i+m\delta \tau)-x(t_i)]^2+[y(t_i+m\delta \tau)-y(t_i)]^2\right)
	\notag
\end{align}
where $x$ and $y$ are the 2D co-ordinates obtained from tracking; and the video contains $N$ frames separated in increments of $\delta t$ seconds. Then a power-law was fit to the time-averaged mean squared displacements using the in scipy.optimize package Python.

\section{Conclusion}
In this paper, we have formulated the space-dependent variable-order fractional master equation \eqref{master} and found its asymptotic solution \eqref{laplace_asympsol2} describing the clustering of particles due to a space dependent fractional exponent. In the continuous limit as the width of subintervals approaches zero, the solution to this master equation converges to that in the space-dependent variable-order fractional diffusion equation found in \cite{fedotov2019asymptotic}. This solution describing the non-uniform distribution of lysosomes inside the cell is confirmed by Monte Carlo simulations of the fractional master equation. We present a new mechanism for lysosome clustering due to a non-uniform fractional exponent. This is a possible explanation for the non-random, stable and yet distinct spatial distribution of lysosomes in the intracellular space found biologically \cite{ba2018whole}. The distribution of lysosome density in living cells matches the asymptotic probability density function \eqref{asympsol} which is the solution of the space-dependent variable-order fractional diffusion and master equation, seen in Figure \ref{fig:lysodata}. Finally, lysosomes have been found to cluster dynamically based on ER spatial density and interact with a variety of other organelles dependent on position and particle density \cite{ba2018whole}. These dynamic interactions can be described in terms of non-linear subdiffusive fractional equations similar to \cite{fedotov2013random,straka2015transport}. Our results give an alternative explanation for clustering and intracellular transport of gold nanoparticles within lysosomes in living cells \cite{liu2017real}. It is important because gold nanoparticles have proven to be promising radiosensitizers for improving proton therapy, since they enhance the radiation damage to tumour cells \cite{lin2015biological,sotiropoulos2017modelling,villagomez2019physical}.

\vskip6pt

\enlargethispage{20pt}

%\ethics{Insert ethics statement here if applicable.}
%
%\dataccess{Insert details of how to access any supporting data here.}

\section{Author Contributions}
SF and AYZ conceived and designed the study and drafted the manuscript. DH conceived and designed the study, drafted the manuscript, carried out the experiments and performed data analysis. MJ and VJA carried out the experiments. All authors read and approved the manuscript.

%\competing{Insert any competing interests here. If you have no competing interests please state 'The author(s) declare that they have no competing interests'.}

%\funding{The authors acknowledge financial support from the EPSRC Grant No. EP/J019526/1, RSF project 20-61-46013 and the Wellcome Trust Grant No. 215189/Z/19/Z.}

\section{Acknowledgments}
SF and AYZ acknowledge financial support from RSF project 20-61-46013. DH acknowledges financial support from the Wellcome Trust Grant No. 215189/Z/19/Z. VJA acknowledges financial support from the EPSRC Grant No. EP/J019526/1.

%\disclaimer{Insert disclaimer text here if applicable.}

%%%%%%%%%% Insert bibliography here %%%%%%%%%%%%%%

\bibliography{real}

\end{document}